\documentclass[twocolumn,showpacs,preprintnumbers,amsmath,amssymb]{revtex4}
\usepackage{graphicx}
\usepackage{dcolumn}
\usepackage{bm}

\newcommand{\be}{\begin{equation}}
\newcommand{\ee}{\end{equation}}
\newcommand{\bea}{\begin{eqnarray}}
\newcommand{\eea}{\end{eqnarray}}
\newcommand{\ba}{\begin{array}}
\newcommand{\ea}{\end{array}}
\newcommand{\nn}{\nonumber \\}
\newcommand{\half}{\frac{1}{2}}

\newcommand{\ovc}{{\overline{c}}}
\newcommand{\calO}{{\cal O}}
\begin{document} 


\title{Triviality of Higher Derivative Theories}

\author{Victor O. Rivelles}
\altaffiliation[On leave from ]{Instituto de F\'{\i}sica, Universidade de
  S\~{a}o Paulo, 
 Caixa Postal 66318, 05315-970, S\~{a}o Paulo, SP, Brazil}
\email{rivelles@lns.mit.edu}
\affiliation{Center for Theoretical Physics \\ Massachusetts Institute of
  Technology \\ Cambridge, MA  02139, USA}

\begin{abstract}
We show that some higher derivative theories have a BRST
symmetry. This symmetry is due to the higher derivative structure and
is not associated to any gauge invariance. If physical states are
defined as those in the BRST cohomology then the only physical state
is the vacuum. All negative norm states, characteristic of higher
derivative theories, are removed from the physical sector. As a
consequence, unitarity is recovered but the S-matrix is trivial. We 
show that a class of higher derivative quantum gravity theories have
this BRST symmetry so that they are consistent as quantum field
theories. Furthermore, this BRST symmetry may be present in both
relativistic and non-relativistic systems. 

\end{abstract}


\maketitle

Higher derivative (HD) theories were introduced in quantum field
theory in an attempt to get rid of ultraviolet divergences
\cite{Thirring}. It was soon recognized that they have an energy which
is not bounded from below \cite{Pais:1950za} 
and that these negative energy states can be traded by negative norm
states (or ghosts) \cite{Heisenberg} leading to non unitary
theories. Although inconsistent, HD theories have better
renormalization properties than conventional ones and a large amount
of work  was dedicated to the study of such theories
\cite{toomany}. In the gravitational context, for instance, it is able
to produce a renormalizable quantum gravity theory
\cite{Stelle:1976gc}. To overcome the ghost problem  
many attempts were done mainly by imposing some superselection rule
or some subsidiary condition to remove the undesirable states
\cite{toomany}. However, there is no general scheme to remove the
ghosts and this is done in a case by case basis. 

In this letter we will show that some HD theories have a BRST
symmetry after ghost fields are introduced. Usually the BRST symmetry
is found in gauge theories as a symmetry of the gauge fixed
action \cite{Rivelles:1995gb}. Its purpose is to remove the negative norm
states associated to the gauge invariance. Physical states
are then defined as those which have zero ghost number and are
invariant under the BRST symmetry. In the present case, however, the
BRST symmetry is not due to gauge invariance. Rather, it is a feature
of the HD structure of the action. As it will be shown, it can be
found even in the case of a HD real scalar field. Since we have a BRST
symmetry it 
is natural to require that the physical states are those which have
zero ghost number and are left invariant by this symmetry, in analogy
to what is done in gauge theories. With these
requirements we find that HD theories have only one physical state,
the vacuum, since all others physical states appear in zero norm
combinations. Therefore, the resulting theory is ghost free but it is
also empty. In this way the unitarity problem associated to HD theories can
be overcome but the resulting theory has no states besides the vacuum.
We will show that this symmetry is present in HD non-Abelian gauge theories
and also in HD gravity theories, so that they are unitary but have a
trivial S-matrix. 

Even though we work mostly with relativistic systems, relativistic
invariance is not a necessary ingredient. The HD BRST symmetry may also
be found in non-relativistic situations. We will exemplify this for
the case of the HD harmonic oscillator. 

Let us first consider a real scalar field $\phi$ in $d$ dimensions
with an action  
\be
\label{1}
S = \int d^dx \left( \half \calO^n \phi \calO^n \phi - \ovc \calO^n c
\right), 
\ee
where $\calO = \prod_{i=1}^N (\Box + m_i^2)$ is a product of $N$
Klein-Gordon operators with masses $m_i$, $c$ is a ghost, $\ovc$ an
anti-ghost and $n$ is an integer. This HD action is invariant under
the following BRST transformations   
\be
\label{2}
\delta \phi = c, \quad \delta c = 0, \quad \delta \ovc = \calO^n 
\phi. 
\ee 
These transformations are nilpotent for $\phi$ and $c$ and on-shell
nilpotent for $\ovc$. To find out the role of this symmetry let us
introduce an auxiliary field $b$ and rewrite the action (\ref{1}) as 
\be
\label{3}
S = \int d^dx \left( b \calO^n \phi - \half b^2 - \ovc \calO^n c
\right).
\ee
The BRST transformations now read 
\be
\label{4}
\delta \phi = c, \quad \delta c = 0, \quad \delta \ovc = b, \quad
\delta b = 0,
\ee
and are off-shell nilpotent. This action can now be written as a total
BRST transformation 
\be
\label{5}
S =  \int d^dx \delta \left[ \ovc ( \calO^n \phi - \half b ) \right],
\ee
indicating that theory may be empty. It looks like a topological  
theory of Witten type \cite{Birmingham:1991ty} but it clearly depends
on the local structure of the space-time. In topological theories
there are no local degrees of freedom from the start while here we
have positive and negative norm states before introducing the ghost
fields. The requirement that the physical states are in the cohomology
of the BRST charge means that all of them, except the vacuum, appear
in zero norm combinations as we will show.  

Interactions can be introduced trough a prepotential
$U(\phi)$. The action 
\be
\label{6}
S = \int d^dx \delta \left[ \ovc ( \calO^n \phi - \half b - U )
\right],
\ee
which is invariant under the BRST transformations (\ref{4}), yields, 
after elimination of the auxiliary field, 
\be
\label{7}
S = \int d^dx \left[ \half (\calO^n \phi)^2 + \half U^2 + 
U \calO^n \phi - \ovc \left( \calO^n c + U^\prime  c
\right) \right].
\ee
Assuming that the BRST symmetry has no anomaly it is enough to examine
the free case to find out the physical states. 

In order to simplify the analysis we will consider from now on 
the case $n=1$ and only one factor in $\calO$, that is $N=1$. The
simplest situation 
is the one dimensional case, that is, quantum mechanics, where $\calO
= d^2/dt^2 + m^2$. Then the action (\ref{1}) reduces to that of a HD
harmonic 
oscillator. Canonical quantization is straightforward if we keep the
auxiliary field and use the action in the form (\ref{3}) where 
only first order derivatives appear after an integration by
parts is performed. We then find that the non trivial equal 
time commutation relations are
\be
\label{8} 
[ \phi, \dot{b} ] = [ b, \dot{\phi} ] = - i, 
\ee
where dots denote time derivatives. The solution to the field equation
for $\phi$ and $b$ can be expanded as 
\bea
\label{9}
\phi(t) &=& \frac{1}{2 m^{3/2}} \left[ (a + b mt) e^{-imt} + (a^\dagger +
    b^\dagger mt ) e^{imt} \right], \quad \\
\label{9a}
b(t) &=& -i m^{1/2} \left( b e^{-imt} - b^\dagger e^{imt} \right). 
\eea
The canonical Hamiltonian yields 
\be
\label{11}
H = m ( 2 b^\dagger b - i b^\dagger a + i a^\dagger b ),
\ee
and using the transformation $\tilde{a} = ia - b/2$, $\tilde{b} = ia
-3b/2$ we find
\be
\label{12}
H = m (\tilde{a}^\dagger \tilde{a} - \tilde{b}^\dagger \tilde{b} ),
\ee
so that it is not positive definite at the classical level. 
Upon quantization we find that the non-trivial commutators for the
operators $a$ and $b$ are given by 
\be
\label{10}
[a, a^\dagger] = -1, \quad [a, b^\dagger ] = i,
\ee
where $a$ and $b$ are annihilation operators and
$a^\dagger$ and $b^\dagger$ are creation ones. 
By the transformation $\hat{a} = b - ia$, $\hat{b} = -ia$, we find
\be
\label{13}
[ \hat{a}, \hat{a}^\dagger ] = - [ \hat{b}, \hat{b}^\dagger ]
= 1,
\ee
so that the Hilbert space is not positive definite. 

We now consider the ghost fields in (\ref{1}). Quantization is
straightforward. The solution to the equations of motion can be
expanded as
\bea
\label{14}
c(t) &=& \frac{1}{2m^{3/2}} \left( c e^{-i\omega t} + c^\dagger
e^{i\omega t} \right), \\ 
\label{15}
\ovc(t) &=& m^{1/2} \left( 
\ovc e^{-i\omega t} + \ovc^\dagger e^{i\omega t} \right),
\eea
where $c$ and $\ovc$ are annihilation operators and $c^\dagger$ and
$\ovc^\dagger$ are creation operators. The non trivial
anti-commutators are given by 
\be
\label{16}
\{c, \ovc^\dagger\} = - \{ \ovc, c^\dagger \} = 1.
\ee
The BRST transformations (\ref{4}) for the operators in (\ref{9}),
(\ref{9a}), (\ref{14}) and (\ref{15}) (not for the fields) are 
\be
\label{17}
\delta a = c, \quad \delta c = 0, \quad \delta \ovc = - i b, \quad
\delta b = 0.
\ee
This means that the these operators belong to the quartet
representation of the BRST algebra \cite{Kugo:gm}. As a consequence,
the BRST 
invariant states, build out of these operators, appear only in zero norm
combinations through the quartet mechanism. Therefore the only 
physical state is the vacuum. In this way the HD harmonic oscillator
is free of negative norm states but the only state left is the vacuum.  

The situation for the scalar field is analogous to the quantum
mechanical case. Now $\calO = \Box + m^2$ and the fields have
expansions similar to (\ref{9}),(\ref{9a}),(\ref{14}) and (\ref{15}) with
the creation and annihilation operators depending on the momenta. They
obey commutation relations similar to (\ref{10}) and (\ref{16}) and
the Hamiltonian has the 
same form as (\ref{11}). Hence the BRST transformations are the same
as in (\ref{17}) and the quartet mechanism is also operational in this
case. The only physical state is the vacuum. The massless case needs
some care but again the same result is found.

At this point it is worth to remind that to remove the negative norm
states for the HD scalar field theory the condition $b=0$ is often
employed \cite{scalar}. It is argued that there is a ``gauge
symmetry''  $\delta \phi = \Lambda$ with $\Lambda$ satisfying $(\Box +
m^2) \Lambda = 0$. The ``gauge condition'' $b=0$ is then imposed to
remove the ghost states. However, the Hamiltonian analysis reveals
that there is no true gauge symmetry since there are no 
constraints at all. Amazingly, a ``BRST symmetry'' associated to this
``gauge symmetry'' can be found \cite{Flato:1986rx} and it agrees with
(\ref{17}). Now its origin is clear; it is completely due to the HD
structure as we have shown.  

We can now easily generalize this construction to other types of
fields. Since we want to keep the same BRST transformations (\ref{4})
then the ghosts and the auxiliary field $b$ must have the same tensor
structure as the field under consideration. For gauge theories we
should also take into account the ordinary Faddeev-Popov ghosts
associated to the gauge symmetry since they have a different origin
from the ghosts coming from the HD structure \cite{Gavrielides:qw}. We
will argue, however, that only one set of ghost fields is needed.

For a theory with a non-Abelian vector field we start with 
\be
\label{20}
S = \int d^dx \, Tr \, \delta \left[ \ovc^\mu ( E_\mu - \half b_\mu )
  \right],
\ee
where $E_\mu$ depends only on the vector field $A_\mu$. The BRST
  transformations are given by 
\be
\label{21}
\delta A_\mu = c_\mu, \quad \delta c_\mu = 0, \quad \delta \ovc_\mu =
b_\mu, \quad \delta b_\mu = 0.
\ee
This structure is similar to that found in topological field theories
\cite{Brooks:1988jm}. There, $E, b$ and $\ovc$ are two
forms instead of vectors so that a topological invariant is generated
after the elimination of the auxiliary field. Here the tensor
structure is quite different and $E_\mu$ is chosen so that it gives 
rise to a HD theory. For a fourth order gauge theory we can choose 
\be
\label{22}
E_\mu = D^\nu F_{\nu\mu} + \frac{1}{2\xi} \partial_\mu \partial^\nu
A_\nu.
\ee
The first term in (\ref{22}) will produce a gauge invariant term in
the action if the second term is absent. The second term breaks gauge
invariance and can be regarded as a gauge fixing term 
since it allows us to find the propagator for $A_\mu$. At this point
we should decide whether we introduce the gauge fixing term in
$E_\mu$, as in (\ref{22}), or if we consider the ordinary
Faddeev-Popov procedure for gauge fixing. If we do not add the gauge
fixing term in (\ref{22}) then $E_\mu$  will give rise to an action 
for $\ovc_\mu$ and $c_\mu$ which has a gauge symmetry  $\delta
\ovc_\mu = D_\mu \overline{\Sigma}, \,\, \delta c_\mu = D_\mu \Sigma$,
where $\overline{\Sigma}$ and $\Sigma$ are 
Grassmannian functions. This would require the introduction of ghosts
for ghosts besides the ordinary Faddeev-Popov ghosts for $A_\mu$. We
then choose to introduce the gauge fixing term directly 
in $E_\mu$ to avoid the proliferation of ghost fields.

With the choice (\ref{22}) we find that (\ref{20}) yields the
following action 
\bea
\label{23}
&&S = \half \int d^dx \, Tr \left(  D^\nu F_{\nu\mu} D_\rho F^{\rho\mu} +
\frac{1}{4\xi^2} \partial^\mu \partial A \partial_\mu \partial A \right. \nn
&& \left. +  \frac{1}{\xi} D^\nu F_{\nu\mu} \partial^\mu \partial A 
+ \overline{{\cal F}}^{\mu\nu} {\cal F}_{\mu\nu} - 2i \{\ovc^\mu, c^\nu\}
  F_{\mu\nu}   
\right. \nn && \left. 
+ \frac{1}{\xi} \partial^\mu \ovc_\mu \partial^\nu c_\nu
\right),
\eea
where $\overline{{\cal F}}^{\mu\nu}$ and ${\cal F}_{\mu\nu}$ are the
field strengths for $\ovc^\mu$ and $c_\mu$, respectively. 
Canonical quantization is easily performed keeping the auxiliary
field $b_\mu$ since it gives rise to a first order Lagrangian. No
constraints are found and the Hamiltonian is not positive
definite. Quantization simplifies in the Feynman gauge $\xi = 1$ and
we find that the quartet mechanism applies to each component of the 
vector field. Again, the only physical state is the vacuum. 
Coupling to ordinary or HD matter is straightforward. However,
only non-minimal couplings arise. Details will be given elsewhere.

Other choices for $E_\mu$ are possible. For instance, 
\be
E_\mu = \frac{1}{m} D^\nu F_{\nu\mu} + \frac{m}{2} A_\mu,
\ee
gives rise to a fourth order massive vector theory which includes the
standard Maxwell term. However, there is no gauge
invariance in this case. 

For gravity the ghosts and the
auxiliary field are second order symmetric tensors. The action is 
\be
\label{23a}
S = \int d^d x \delta \left[ \ovc^{\mu\nu} ( E_{\mu\nu} - \half
 b_{\mu\nu} ) \right],
\ee
and the BRST transformations are given by
\be
\label{23b}
\delta g_{\mu\nu} = c_{\mu\nu}, \quad \delta c_{\mu\nu} = 0, \quad
 \delta \ovc^{\mu\nu} = b^{\mu\nu}, \quad \delta b^{\mu\nu} = 0.
\ee
For a generic fourth order gravity theory we can choose  
\be
\label{24}
E_{\mu\nu} = g^{1/4} \left( c_1 R_{\mu\nu} + c_2 g_{\mu\nu} R + c_3
g_{\mu\nu} + \dots \right),
\ee
where $g_{\mu\nu}$ is the metric, $g$ its determinant, $R_{\mu\nu}$
the Ricci tensor, $R$ the curvature scalar and dots denote the gauge
fixing terms. After the elimination of the auxiliary field we find
that the action for the gravitational sector can be
written as  
\be
\label{24a}
S_{gr} = \int d^dx \, \sqrt{-g} \left[ \gamma R - \Lambda + \alpha R^2
  + \beta R^{\mu\nu} R_{\mu\nu} \right], 
\ee
with $\gamma^2 = - 4 \Lambda ( d\alpha + \beta)/d$. Gauge fixing terms
were omitted. Notice that we can not get the Einstein-Hilbert action
since 
setting $\alpha = \beta = 0$ to get rid of the HD contributions also
eliminates the term in $R$. However, we can choose $\Lambda =
0$ in order to get a purely HD gravity theory. Alternatively we can
choose $\beta = -d\alpha$, which corresponds to a traceless Ricci
tensor. This case is distinguished by energy considerations
\cite{Deser:2002rt}. The conformal case has no special features. 
Notice also that (\ref{24}) does not allow us to
generate the square of the Riemann tensor thus precluding Gauss-Bonnet
terms which arise in string theory. 
Details will be presented
elsewhere.

Some final remarks are in order. 
Could a BRST symmetry be found for ordinary second order
theories? Covariance requires that the operator $\calO$ be at least
quadratic in the derivatives. However, in lower dimensions, this
requirement can be relaxed. Chiral bosons in two dimensions have a
BRST symmetry which allows the vacuum as the only physical state
\cite{Girotti:1992hy}. In higher dimensions we could consider the
action (\ref{6}) without the term $\calO^n \phi$ and $U$ as a function
of $\phi$ and its derivatives. Then we can take $U=\sqrt{\phi(\Box + m^2)
  \phi}$. The term $U^2$ in (\ref{7}) yields the 
Lagrangian for the ordinary scalar field but we get non-local
interactions with the ghosts due to the term $U^\prime \ovc
c$. Even though there is a BRST symmetry, a detailed analysis shows
that on-shell $\ovc = b = 0$ so that the quartet 
mechanism can not be applied. It is imperative to have non-trivial
solutions to the field equation for the quartet mechanism to
work. Since $\ovc=0$ all 
contributions from the ghost sector to the physical Hilbert space
vanish because they must be functions of $\ovc c$. Then we end up with
the usual scalar theory and the BRST symmetry is trivial in this case.

A more radical possibility would be to consider non-local expression
for $\calO$, for instance $\calO = \sqrt{\Box +
m^2}$. The action (\ref{1}) with $n=1$ would be the ordinary one for a
scalar field but for the ghosts we find a non-local
action. Quantization of such non-local theories has been done 
\cite{Barci:1995ad} but it not clear how the quartet mechanism can be
applied with non-local ghosts. This deserves further understanding. 

Returning to the quantum mechanical case, 
the action (\ref{7}) resembles that of supersymmetric quantum
mechanics if $\calO = d/dt$ and the ghosts are regarded as fermions. The 
BRST transformations (\ref{4}) are identical to the supersymmetry
transformations but the anti-BRST transformation (obtained by
replacing ghosts by anti-ghosts and vice-versa) 
$\overline{\delta} \phi = \ovc, \quad \overline{\delta} \ovc = 0, \quad
\overline{\delta} c = - ( \calO + V ),$ are not. The anti-commutator
of the BRST and anti-BRST transformations 
vanishes as expected while in the supersymmetric case it would be
proportional to the Hamiltonian. There are crucial signs in the action
and in the transformation rules which, when changed, produces the
supersymmetric model. Hence the supersymmetric model is obtained by
twisting the BRST symmetry. 

We should also remark that the BRST symmetry can not be found
in any HD theory. For instance, in the simple case of non-degenerated
masses 
\be
\label{27}
S = \int d^dx \left[ \half (\Box + m_1^2) \phi (\Box + m_2^2) \phi - \ovc
(\Box + m_1^2) c \right],
\ee
there is a symmetry of the action given by
$\delta \phi = c, \quad \delta c = 0, \quad \delta \ovc = (\Box +
m_2^2) \phi.$ However, this symmetry is not nilpotent since on-shell
$\delta^2 \ovc 
= (m_2^2 - m_1^2) c \not= 0$. It is nilpotent only when both masses are
equal thus implying in (\ref{1}) with $n=1$ and $\calO = \Box + m^2$, 
where $m$ is the common mass. 

As a last remark it must be stressed that a trivial topology for the
space (or space-time) was assumed throughout the paper. If a
non-trivial topology is present then topological excitations may arise
as they do in topological field theories of Witten type. For instance,
in the case of non-Abelian gauge fields (\ref{23}) there are instanton
solutions which may give non-trivial contributions to correlation
functions. This is presently under investigation and will be reported
elsewhere.

\begin{acknowledgments}
I would like to thank A. J. Accioly, S. Deser, R. Jackiw, I. Shapiro
and S. Sorella for useful discussions. 
This work was partially supported by CAPES, CNPQ and PRONEX under contract
CNPq 66.2002/1998-99. 
\end{acknowledgments}


\begin{thebibliography}{100}

\bibitem{Thirring}
W. Thirring, 
Phys. Rev. {\bf 77}, 570 (1950).

\bibitem{Pais:1950za}
A.~Pais and G.~E.~Uhlenbeck,
Phys.\ Rev.\  {\bf 79}, 145 (1950).

\bibitem{Heisenberg}
W.~Heisenberg, 
Nucl.\ Phys.\ {\bf4}, 532 (1957).

\bibitem{toomany} See the following papers for earlier references: 
J. Lukierski, Acta Phys. Polon. {\bf 32}, 771 (1967);
N.~Nakanishi,
Prog.\ Theor.\ Phys.\ Suppl.\  {\bf 51}, 1 (1972);
K.~S.~Stelle,
Gen.\ Rel.\ Grav.\  {\bf 9}, 353 (1978);
H.~Narnhofer and W.~Thirring,
Phys.\ Lett.\  {\bf 76B}, 428 (1978);
M.~Mintchev,
J.\ Phys.\ A {\bf 13} (1980) 1841;
E.~S.~Fradkin and A.~A.~Tseytlin,
Nucl.\ Phys.\ B {\bf 201}, 469 (1982);
P.~Broadbridge,
J.\ Phys.\ A {\bf 16}, 3271 (1983);
N.~H.~Barth and S.~M.~Christensen,
Phys.\ Rev.\ D {\bf 28}, 1876 (1983);
U.~Moschella,
J.\ Math.\ Phys.\  {\bf 31}, 2480 (1990);
I.~L.~Buchbinder, S.~D.~Odintsov and I.~L.~Shapiro,
``Effective Action In Quantum Gravity,''(IOP, Bristol, 1992);
M.~Asorey, J.~L.~Lopez and I.~L.~Shapiro,
Int.\ J.\ Mod.\ Phys.\ A {\bf 12}, 5711 (1997)
[arXiv:hep-th/9610006];
F.~J.~de Urries, J.~Julve and E.~J.~Sanchez,
J.\ Phys.\ A {\bf 34}, 8919 (2001)
[arXiv:hep-th/0105301];
S.~W.~Hawking and T.~Hertog,
Phys.\ Rev.\ D {\bf 65}, 103515 (2002)
[arXiv:hep-th/0107088];
T.~C.~Cheng, P.~M.~Ho and M.~C.~Yeh,
Nucl.\ Phys.\ B {\bf 625}, 151 (2002)
[arXiv:hep-th/0111160];
A.~Accioly, A.~Azeredo and H.~Mukai,
J.\ Math.\ Phys.\  {\bf 43}, 473 (2002);
E.~J.~Villasenor,
J.\ Phys.\ A {\bf 35}, 6169 (2002)
[arXiv:hep-th/0203197];
S.~De Filippo and F.~Maimone,
AIP Conf.\ Proc.\  {\bf 643}, 373 (2003)
[arXiv:gr-qc/0207059];
G.~Magnano and L.~M.~Sokolowski,
arXiv:gr-qc/0209022;
S.~Nojiri,
arXiv:hep-th/0210056.

\bibitem{Stelle:1976gc}
K.~S.~Stelle,
Phys.\ Rev.\ D {\bf 16}, 953 (1977).

\bibitem{Rivelles:1995gb} The BRST transformations may be written in
  different ways (even in non local form) but that is not the case
  here. This is discussed in 
V.~O.~Rivelles,
Phys.\ Rev.\ Lett.\  {\bf 75}, 4150 (1995)
[arXiv:hep-th/9509028];
V.~O.~Rivelles,
Phys.\ Rev.\ D {\bf 53}, 3247 (1996)
[arXiv:hep-th/9510136];
V.~O.~Rivelles,
Class.\ Quant.\ Grav.\  {\bf 19}, 2525 (2002)
[arXiv:hep-th/0109171].

\bibitem{Birmingham:1991ty}
D.~Birmingham, M.~Blau, M.~Rakowski and G.~Thompson,
Phys.\ Rept.\  {\bf 209}, 129 (1991).

\bibitem{Kugo:gm}
T.~Kugo and I.~Ojima,
Prog.\ Theor.\ Phys.\ Suppl.\  {\bf 66}, 1 (1979).

\bibitem{scalar}
D.~Zwanziger,
Phys.\ Rev.\ D {\bf 17}, 457 (1978);
A.~Z.~Capri, G.~Grubl and R.~Kobes,
Annals Phys.\  {\bf 147}, 140 (1983);
U.~Moschella and F.~Strocchi,
Lett.\ Math.\ Phys.\  {\bf 19} (1990) 143; 
U.~Moschella and F.~Strocchi,
Lett.\ Math.\ Phys.\  {\bf 19} (1990) 143.

\bibitem{Flato:1986rx} In fact, it was found in the singleton field
  theory in 
M.~Flato and C.~Fronsdal,
Phys.\ Lett.\ B {\bf 189}, 145 (1987).

\bibitem{Gavrielides:qw}
A.~Gavrielides, T.~K.~Kuo and S.~Y.~Lee,
Phys.\ Rev.\ D {\bf 13}, 2912 (1976).

\bibitem{Brooks:1988jm}
R.~Brooks, D.~Montano and J.~Sonnenschein,
Phys.\ Lett.\ B {\bf 214}, 91 (1988).

\bibitem{Deser:2002rt}
S.~Deser and B.~Tekin,
Phys.\ Rev.\ Lett.\  {\bf 89}, 101101 (2002)
[arXiv:hep-th/0205318];
S.~Deser and B.~Tekin,
arXiv:hep-th/0212292.

\bibitem{Girotti:1992hy} 
H.~O.~Girotti, M.~Gomes and V.~O.~Rivelles,
Phys.\ Rev.\ D {\bf 45}, 3329 (1992)
[arXiv:hep-th/9202043].

\bibitem{Barci:1995ad}
D.~G.~Barci, L.~E.~Oxman and M.~Rocca,
Int.\ J.\ Mod.\ Phys.\ A {\bf 11}, 2111 (1996)
[arXiv:hep-th/9503101];
D.~G.~Barci and L.~E.~Oxman,
Mod.\ Phys.\ Lett.\ A {\bf 12}, 493 (1997)
[arXiv:hep-th/9611147].


\end{thebibliography}
\end{document}